\documentstyle[preprint,aps]{revtex}
\input{psfig}
\begin{document}
\tightenlines
\draft
\preprint{
}
\title{Sideward flow in Au + Au collisions between 2$A$ and 8$A$ GeV}

\author{
H.~Liu,$^7$ 
N.N.~Ajitanand,$^{12}$ J.~Alexander,$^{12}$ M.~Anderson,$^5$ 
D.~Best,$^1$ F.P.~Brady,$^5$ T.~Case,$^1$ W.~Caskey,$^5$ 
D.~Cebra,$^5$ J.~Chance,$^5$ 
B.~Cole,$^4$ K.~Crowe,$^1$ A.~Das,$^{10}$
J.~Draper,$^5$ M.~Gilkes,$^{12}$ S.~Gushue,$^2$
M.~Heffner,$^5$ A.~Hirsch,$^{11}$
E.~Hjort,$^{11}$ L.~Huo,$^6$ M.~Justice,$^7$
M.~Kaplan,$^3$ D.~Keane,$^7$ J. Kintner,$^8$
J.~Klay,$^5$ D.~Krofcheck,$^9$ R.~Lacey,$^{12}$ M.A.~Lisa,$^{10}$
Y.M.~Liu,$^6$ R.~McGrath,$^{12}$ Z.~Milosevich,$^3$ G.~Odyniec,$^1$
D.~Olson,$^1$ S.Y.~Panitkin,$^7$ 
N.~Porile,$^{11}$ G.~Rai,$^1$ H.G.~Ritter,$^1$
J.~Romero,$^5$ R.~Scharenberg,$^{11}$ L.S.~Schroeder,$^1$
B.~Srivastava,$^{11}$ N.T.B.~Stone,$^1$ T.J.M.~Symons,$^1$ 
S.~Wang,$^7$ J.~Whitfield,$^3$ T.~Wienold,$^1$ R. Witt,$^7$ 
L.~Wood,$^5$ X.~Yang,$^4$ W.N.~Zhang,$^6$ Y.~Zhang$^4$
\\
(E895 Collaboration)
}
\address{
$^1$Lawrence Berkeley National Laboratory, Berkeley, California 94720\\
$^2$Brookhaven National Laboratory, Upton, New York 11973\\
$^3$Carnegie Mellon University, Pittsburgh, Pennsylvania 15213\\
$^4$Columbia University, New York, New York 10027\\
$^5$University of California, Davis, California 95616\\
$^6$Harbin Institute of Technology, Harbin 150001, P.~R.~China \\
$^7$Kent State University, Kent, Ohio 44242\\
$^8$St.~Mary's College of California, Moraga, California 94575\\
$^9$University of Auckland, Auckland, New Zealand \\
$^{10}$The Ohio State University, Columbus, Ohio 43210\\
$^{11}$Purdue University, West Lafayette, Indiana 47907\\
$^{12}$State University of New York, Stony Brook, New York 11794\\
}

\maketitle

\begin{abstract}
Using the large acceptance Time Projection Chamber of experiment E895 at 
Brookhaven, measurements of collective sideward flow in Au + Au collisions 
at beam energies of 2, 4, 6 and 8$A$ GeV are presented in the form of 
in-plane transverse momentum $\langle p_x \rangle$ and the first 
Fourier coefficient of azimuthal anisotropy $v_1$.  These measurements   
indicate a smooth variation of sideward flow as a function of beam energy.  
The data are compared with four nuclear transport models which have an 
orientation towards this energy range.  All four exhibit some qualitative 
trends similar to those found in the data, although none shows a consistent 
pattern of agreement within experimental uncertainties.  
\end{abstract}
\narrowtext

Sideward flow was the first type of collective motion to be identified 
among fragments from relativistic nuclear collisions \cite{HGRreview97}.  
It consists of a preferential emission in the plane defined by the 
incident nuclei (the reaction plane); at relativistic energies, nucleon 
emission towards the projectile side is favored forward of the center of 
mass rapidity, while the target side is favored at backward rapidities.  
This behavior is normally attributed to a release of compressional energy, 
and thus is sensitive to the integrated effect of the nuclear pressure 
generated in the collision.  Models indicate that sideward flow is 
established during the early, high density stage of the heavy ion 
collision, and that it is minimally distorted during the subsequent 
evolution.   

A Quark Gluon Plasma (QGP) might be formed in heavy ion collisions at 
sufficiently high energies, in contrast to the purely hadronic matter 
that exists throughout the collision process at lower energies.  Near 
the transition between these two regimes, it is argued that the increased 
entropy density leads to a ``softest point" in the nuclear equation of 
state (EOS), which could generate a minimum in the pressure-driven sideward 
flow at the relevant beam energy \cite{Shuryak95,Rischke96,Brachmann99}.  
Earlier, it had been suggested that sideward flow at CERN energy would 
have a magnitude that depends on whether or not a plasma is produced 
\cite{Amelin91}.  Initial calculations incorporating a softest point 
in the framework of one-fluid relativistic hydrodynamics featured a 
prominent minimum in the sideward flow for near-central Au + Au 
collisions around 5$A$ GeV \cite{Rischke96}.  However, the softening 
effect is reduced when allowance is made for the finite size of the 
hydrodynamic system \cite{Spieles98}, and it occurs at higher beam 
energies in a three-fluid model \cite{Brachmann99}.  Most recently, it  
has been argued that the shape of the rapidity dependence of sideward 
flow may be a QGP signature \cite{Csernai99}.  

In the past, hydrodynamic calculations have frequently been the first 
to predict new collective phenomena, whereas microscopic transport 
models have typically reproduced flow measurements subsequently  
with better agreement.  Relativistic transport codes yield measurably 
different flow amplitude near the ``softest point" beam energy, 
depending on whether or not a phase transition is simulated 
\cite{Li98,Pawel98}.  Furthermore, the smooth excitation function for 
elliptic flow observed in E895 \cite{Pinkenburg98} has been interpreted 
as a possible phase transition signature because a transport model 
comparison is consistent with a progressive softening of the EOS with 
increasing beam energy \cite{Pawel98}.  However, a transition from 
hadronic to string degrees of freedom has since been put forward as an 
alternative interpretation for such an EOS softening \cite{Sahu99}.  
There are several further reasons why the 2-8$A$ GeV energy range is 
especially interesting: it is largely unexplored; many inelastic NN 
channels open up within this relatively narrow range of beam energies; 
and models suggest that the highest baryon density is reached in this 
region \cite{Pang92}.  

We report proton sideward flow measurements for Au + Au collisions at 
kinetic energies of 1.85, 3.9, 5.9, and 7.9$A$ GeV in experiment E895 
\cite{Rai90} at Brookhaven's AGS.  The data presented come from the main 
E895 subsystem --- the EOS Time Projection Chamber (TPC).  E895 
allows a seamless extension to higher beam energies of the detailed flow 
excitation functions already measured \cite{Wang95,Partlan95} using the 
same TPC at the Bevalac.  The TPC offers good acceptance for charged 
particles over a substantial fraction of $4\pi$ solid angle, as well as 
particle identification via energy loss measurement, although ambiguities 
in separating protons from both pions and deuterons increase with beam 
energy.  The full event reconstruction capability of E895 allows 
determination of flow even when the correlations used to establish the 
reaction plane orientation are relatively weak. 

The estimated reaction plane azimuth $\Phi$ for an event is based on 
the orientation of  
${\bf Q} = \sum_j w\, {\bf p}^\perp_j/ p^\perp_j$,  where 
$j$ runs over all baryonic fragments in the event, 
${\bf p}^\perp$ is momentum in the plane perpendicular to the projectile 
direction, and we use the weighting factor 
$w = y_j' / {\rm max}(|y_j'|,~0.8)$, 
where $y_j' = y^{\rm lab}_j/y^{\rm mid} - 1$ is the center-of-mass frame 
rapidity for fragment $j$, and $y^{\rm mid}$ is half the rapidity gap 
between target and projectile \cite{Dan85}.  Thus, $y'$ denotes normalized 
rapidity such that the target and projectile are always at $y' = -1$ and 
$+1$, respectively.  Detector asymmetries and inefficiencies result in 
anisotropies in the determined reaction plane azimuth $\Phi$.  Uniform 
distributions in $\Phi$ are recovered after each track is assigned a 
weight according to its $y$ and ${\bf p}^\perp$ magnitude and direction.  

The centrality of collisions is characterized in terms of charged
ejectile multiplicity $M$ as a fraction of $M_{\rm max}$, the value 
near the upper limit of the $M$ spectrum where the height of the 
distribution has fallen to half its plateau value \cite{Gustafsson84}.  
The events used in our analysis come from the region where sideward 
flow is at or near its maximum ---  multiplicities
between 0.5 and 0.75 times $M_{\rm max}$.  Models indicate that this
region corresponds to impact parameters distributed mostly between 5
and 7 fm.  After centrality selection, the analyzed samples contain 
12, 24, 9.3 and 7.2 thousand events at 2, 4, 6 and 8$A$ GeV, respectively.

\begin{figure}[c]
\centerline{\psfig{file=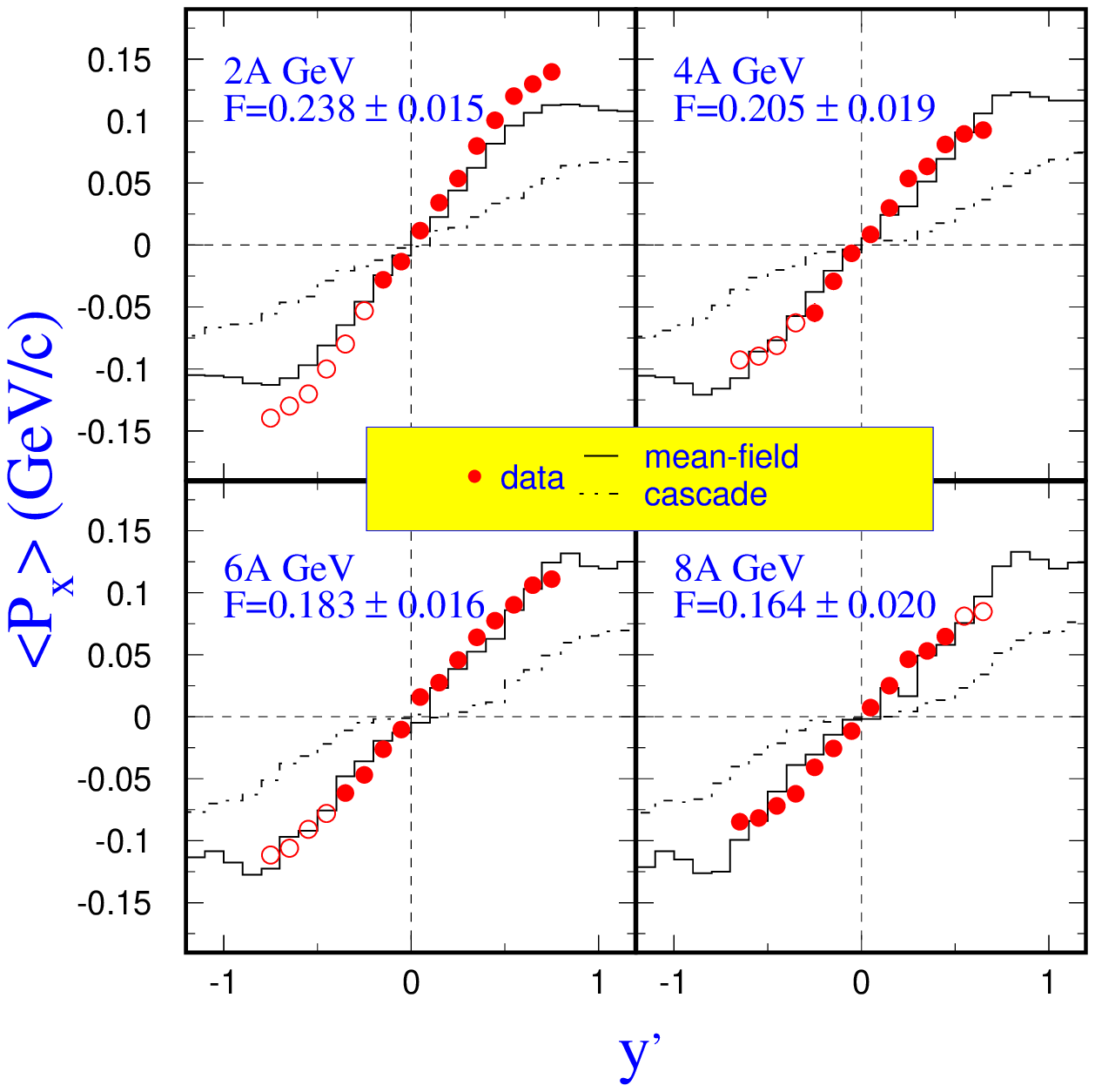,width=13.0cm}}
\vspace{0.2in}
\caption{Average proton $p_x$ as a function of normalized rapidity, 
$y'$ (the target and projectile are always at $y' = -1$ and $+1$, 
respectively).  The
closed symbols are direct measurements and the open symbols are
generated by reflection about midrapidity.  Histograms are RQMD
calculations in cascade mode (dashed line) and mean field mode (solid 
line).}
\label{fig1}
\end{figure}

The mean proton transverse momentum projected onto the reaction plane,
$\langle p_x \rangle$, is presented as a function of rapidity in
Fig.~\ref{fig1}. Using the prescription described in Ref.~\cite{Posk98}, 
these and all subsequent flow signals are corrected for finite resolution 
in determining the reaction plane.  Dispersion correction factors are 
0.89, 0.79, 0.62 and 0.43 at 2, 4, 6 and 8$A$ GeV, respectively.  In E895, 
there are known distortions below $p_\perp \sim 0.3$ GeV$/c$ caused by 
track reconstruction inefficiencies and by the breakdown of proton -- 
$\pi^+$ separation at some rapidities.  At Bevalac beam energies, where 
these distortions are not a factor, we have studied the contours on 
scatter plots of the $p_\perp$ components $p_x$ and $p_y$ within various 
rapidity gates, after events are rotated so that estimated reaction planes 
are aligned with the $x$ axis.  It is observed that the contours are 
concentric, and $\langle p_x \rangle$ is constant in any slice of 
$p_y$, with deviations of less than 5\%.  Therefore, full acceptance in 
the $p_\perp$ plane is not required to extract the sideward flow, provided 
an appropriate $p_y$ cut is applied to remove the problematic region.  For 
$p_y \agt 0.3$ GeV$/c$, the expected flat behavior in $\langle p_x(p_y) 
\rangle$ is observed at E895 beam energies.  We assign these plateau values 
to $\langle p_x \rangle$, with the assumption that the behavior described 
above is a general property of sideward flow which does not change 
between Bevalac and AGS energies.  Moreover, GEANT-based simulations of 
the detector response without this assumption yield corrected $\langle 
p_x \rangle$ results that are consistent within the reported 
uncertainties. 

\begin{figure}[c]
\centerline{\psfig{file=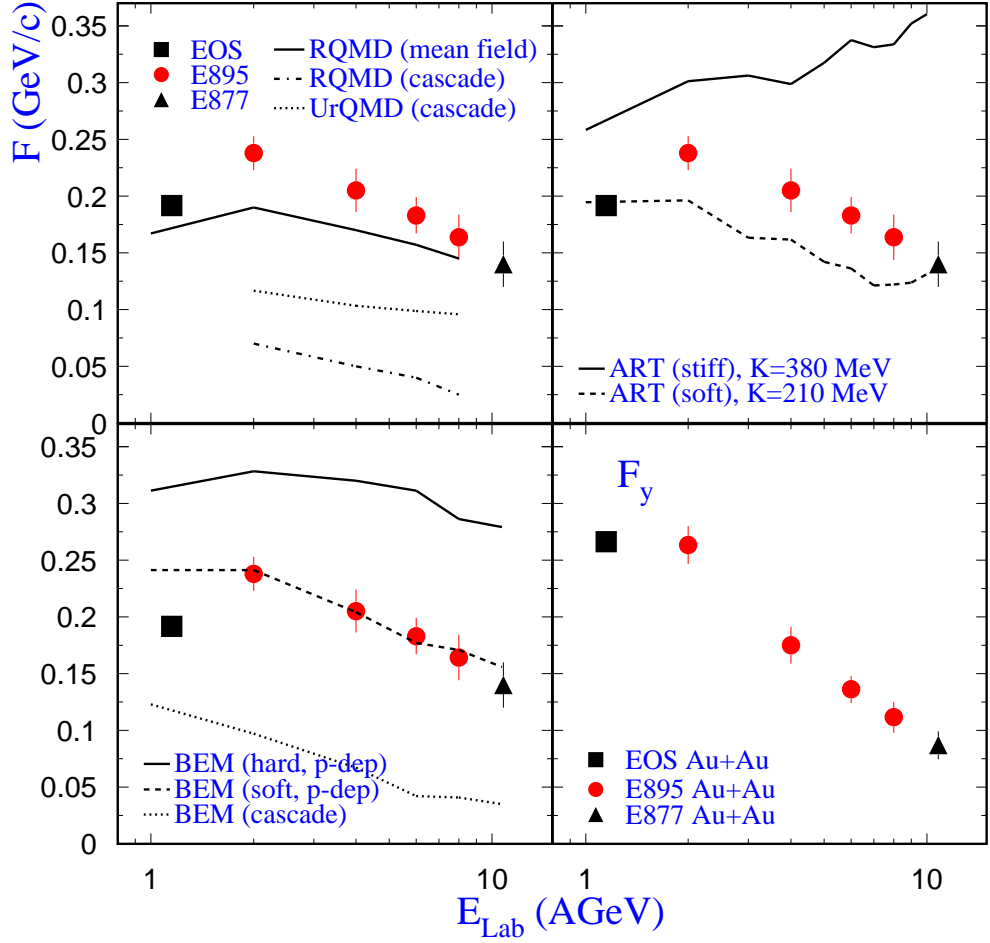,width=13.0cm}}
\vspace{0.2in}
\caption{ Proton flow magnitude as a function of beam energy; the lower 
right panel shows the measured $F_y$, while the other three panels show 
identical measurements of the parameter $F$, with different transport 
model calculations superimposed.  The error bars include systematic 
uncertainties.}
\label{fig2}
\end{figure}

Shapes of $\langle p_x (y)\rangle$ are normally close to linear over 
an interval centered on midrapidity, and a function $F y' + C y'^3$  
typically fits the $\langle p_x (y')\rangle$ distribution over the 
$y'$ region dominated by participant fragments.  It has become common 
to use the fitted linear coefficient $F$ (or $F_y = F/ y^{\rm mid}$, the 
corresponding slope for unnormalized rapidity) to characterize the overall 
strength of the sideward flow.  We average the fitted coefficient $F$ 
with and without imposing $C = 0$, and the difference generally dominates 
the systematic uncertainty in the slope (which is large 
compared with the statistical error).  Figure~\ref{fig2} presents both 
$F$ and $F_y$ as functions of beam energy, along with the same quantities 
for comparable centrality, as measured in the same detector at lower 
energy \cite{Partlan95}, and in E877 \cite{E877} at maximum AGS energy.  
Nucleon rapidity spectra $dN/dy'$ at fixed centrality have close to the 
same form at different beam energies, and this observation motivated 
the use of normalized rapidity $y'$ in the definition of $F$.  Both 
flow observables decrease steadily with increasing beam energy over the 
E895 range.  The significant decrease in $F_y$ at E895 energies (in 
contrast to a flat or slightly decreasing trend in $F_y(E_{\rm beam})$ 
at Bevalac energies \cite{Partlan95}), is interpreted in the hydrodynamic 
picture as an increased deviation from ideal fluid behavior (constant 
$F_y(E_{\rm beam})$ \cite{Balazs84,HGRreview97}), most plausibly 
viscosity arising from the increasingly copious particle production in 
this region \cite{Balazs84}.  Phenomenologically, the observed 
trends in $F$ and $F_y$ between 0.2 and 10$A$ GeV can be related to the 
steady decrease in the azimuthal anisotropy of the proton distribution 
$v_1$ (defined below) over this range, 
in combination with the variation of $\langle p_\perp \rangle$, which 
increases steeply at Bevalac/SIS energies but appears to approach 
saturation at higher beam energies \cite{HGRreview97}. 

\begin{figure}[c]
\centerline{\psfig{file=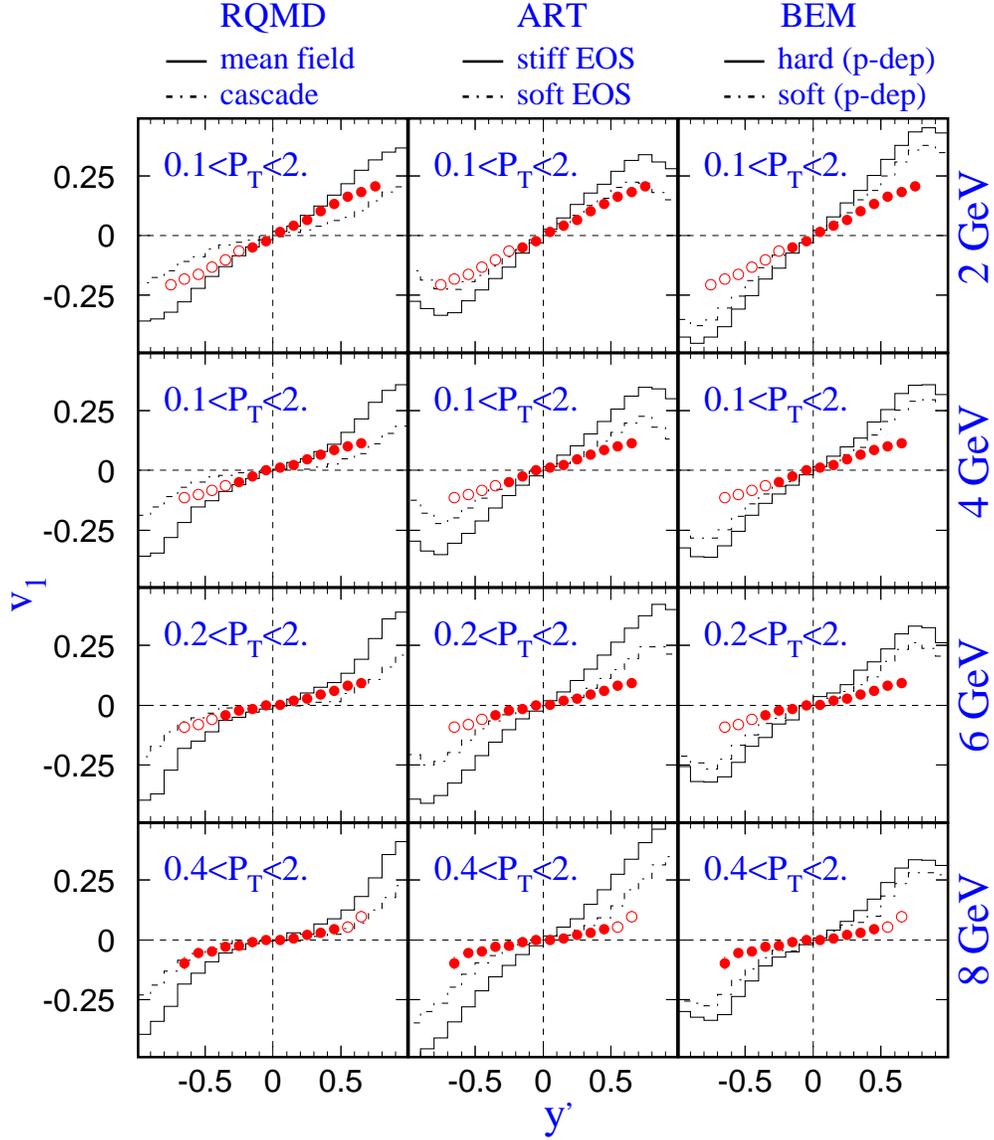,width=13.0cm}}
\vspace{0.2in}
\caption{Fourier coefficients $v_1$ as a function of normalized
rapidity, $y'$.  The closed symbols are direct measurements and the open
symbols are generated by reflection about midrapidity.  The labeled 
transverse momentum gates are in units of GeV/$c$.  Histograms are
transport model calculations, as labeled at the top of each column, 
and each column of four panels shows identical measurements.}
\label{fig3}
\end{figure}

Within a rapidity window, flow causes anisotropic distributions 
of track azimuths $\phi$ relative to the reaction plane.   
These anisotropies generally can be well described by the truncated 
Fourier expansion
\begin{equation}
dN/d\phi \approx v_0 \left[ 1 + 2 v_1\cos\phi + 2 v_2\cos 2\phi \right]\,.
\label{eq:v1v2}
\end{equation}
The first Fourier coefficient, $v_1$, reflects the azimuthal angular 
part of the sideward flow correlation, and is related to the $\langle 
p_x\rangle$ sideward flow observable according to
\begin{equation}
\langle p_x \rangle = \frac{1}{N} \int v_1(p^\perp) p^\perp
                                       \frac{dN}{dp^\perp} dp^\perp\,.
\end{equation}
The $v_2$ coefficient in Eq.~(\ref{eq:v1v2}) represents elliptic flow 
\cite{HGRreview97}, already reported for the present E895 data set 
\cite{Pinkenburg98}.  Figure~\ref{fig3} presents measured $v_1$ 
coefficients for protons as a function of rapidity, at the four E895 
beam energies.  $p_\perp$ gates, as labeled, were applied when generating 
the $v_1(y)$ spectra reported in Fig.~\ref{fig3} (but $p_\perp$ gates 
were not used in any of the $\langle p_x \rangle$ analyses).  

The data in Figs.~\ref{fig1} through \ref{fig3} do not show evidence of a
dip in the flow excitation function \cite{Rischke96,Spieles98,Li98,Pawel98},   
and previous measurements at both higher and lower beam energy are 
consistent with a smooth extrapolation of the E895 data.  The $\langle 
p_x (y)\rangle$ slopes show no evidence of a decrease near midrapidity, 
while a small flattening effect \cite{Liu98,Csernai99} appears in $v_1(y)$ 
at the highest energies.  The decrease in $v_1(y)$ slope at midrapidity is 
very prominent at CERN energy \cite{NA49}.  

Nuclear transport models describe the nuclear collision in part as 
successive point-like nucleon-nucleon interactions. The NN cross sections 
are mostly taken from experiment and include inelastic processes, e.g.,
production of resonances, pions, etc.  These models have been successful
in reproducing a large fraction of the published flow measurements to
within a few tens of percent or better.  The representative transport
models now available include RQMD \cite{Sorge95} (Relativistic Quantum
Molecular Dynamics), UrQMD \cite{Bass98} (Ultrarelativistic Quantum 
Molecular Dynamics), ART \cite{Li98} (A Relativistic Transport model) 
and BEM \cite{Pawel98} (Boltzmann-Equation Model).  In one operating 
mode of RQMD (``cascade''), the hadrons and resonances propagate freely 
between binary collisions, and the equilibrium pressure is close to that 
of an ideal gas.  RQMD also contains an option (the ``mean-field'' mode) 
which allows additional pressure to be generated in the high density 
stage.  UrQMD provides only a cascade mode at these energies, but 
features a completely independent implementation from RQMD.  BEM is based 
on the relativistic Landau theory of quasiparticles.  In addition to a 
cascade mode, two types of momentum-dependent EOS can be selected in BEM: 
a scalar potential (soft EOS, $K = 210$ MeV) and a vector potential (hard 
EOS, $K = 380$ MeV).  In the ART model, there is likewise provision for a 
soft and stiff EOS, but using different phenomenological prescriptions 
from those in BEM.  An early version of the BEM code was used in 
refs~\cite{Pinkenburg98,Rai99}; the version reported here yields less 
sideward flow than before \cite{Rai99}, but almost the same elliptic 
flow \cite{Pinkenburg98,Rai99}.  All 
comparisons presented here are subject to a systematic uncertainty 
arising from the fact that all of the model calculations neglect 
formation of composite nuclear fragments, while the flow measurements 
are for free protons only.  

Although a relatively complex pattern of disagreement is observed between 
data and all four models, several general conclusions are suggested by 
the comparisons.  The cascade modes of BEM, RQMD and UrQMD all exhibit 
less proton flow $F$ than observed.  The significant differences among 
the codes in cascade mode (up to a factor of 2 or more) indicates that 
the binary scattering part of transport simulations, usually considered 
to be better understood than the ``long-range'' part, remains a source 
of significant uncertainty in model calculations at these energies.  The 
soft EOS in ART and BEM, and RQMD's mean field, all come close to 
reproducing the $F$ measurements.  The relative variation of $v_1$ for 
data and all models as a function of increasing beam energy is 
suggestive of a softening trend \cite{Pinkenburg98}.  However, there is 
a marked tendency for the $v_1$ data to favor 
substantially softer equations of state than $F$ measurements.  The 
coefficient $v_1$ reflects only the azimuthal angular part of the sideward 
flow correlation, while $\langle p_x \rangle$ and $F$ also include the 
effect of $p_\perp$ magnitude correlations; thus, the tendency noted 
above indicates that the models consistently have too small a $p_\perp$ 
magnitude flow correlation relative to their azimuthal angle flow 
correlation.  It is also noteworthy that no mode of any of these models 
is close to simultaneously reproducing the E895 elliptic flow  
\cite{Pinkenburg98,Rai99} and our data for $F$ and $v_1$.  

In summary, we report measurements of sideward flow in collisions of heavy 
nuclei in the previously unexplored region between maximum Bevalac/SIS 
energy and maximum AGS energy.  Sideward flow decreases smoothly over the 
2$A$ to 8$A$ GeV range, and extrapolations are consistent with existing 
measurements at both lower and higher beam energies.  A new trend of 
decreasing flow sets-in near the low end of the studied beam energy range 
--- the $F_y$ excitation function changes from flat or slowly decreasing 
to a steeper rate of decrease, while the slope of the $F$ excitation 
function changes sign.  This change roughly coincides with the onset of 
copious particle production, and is reproduced qualitatively by some 
transport calculations.  As at other energies \cite{HGRreview97}, transport 
simulations in cascade mode consistently yield less flow than is observed.  
The model flow calculations for E895 energies are subject to significant 
systematic uncertainties, as inferred from variations among different 
models and from the overall extent and pattern of agreement with experiment.  
At present, these uncertainties appear to be larger than the magnitude of 
flow signatures of physics importance, and so strongly motivate further 
transport model investigation and development focused specifically on this 
unique energy domain where many new inelastic NN channels open up, and meson 
yields increase steeply.  Overall, our flow excitation function measurements 
offer important constraints on possible conclusions regarding a QGP phase 
transition at or above E895 energies. 

We thank the authors of refs.~\cite{Li98,Pawel98,Sorge95,Bass98} for 
providing model codes and/or calculations.  We also thank P.~Danielewicz, 
A.~Poskanzer and S.~Voloshin for valuable input and discussions.  We 
acknowledge DOE, NSF and other funding as detailed in ref.~\cite{Pinkenburg98}, 
and computational resources provided by the National Energy Research Scientific 
Computing Center.  



\end{document}